\documentclass[
]{ceurart}

\usepackage{listings}
\begin{document}

\copyrightyear{2022}
\copyrightclause{Copyright for this paper by its authors.
  Use permitted under Creative Commons License Attribution 4.0
  International (CC BY 4.0).}

\conference{SC$^2$ 2021: 6th International Workshop on Satisfiability Checking and Symbolic Computation, August 19-20 2021}

\title{SC-Square:  Future Progress with Machine Learning?}

\author{Matthew England}[%
orcid=0000-0001-5729-3420,
email=Matthew.England@coventry.ac.uk,
url=https://matthewengland.coventry.domains,
]
\address{Coventry University, Coventry, UK}

\begin{abstract}
The algorithms employed by our communities are often underspecified, and thus have multiple implementation choices, which do not effect the correctness of the output, but do impact the efficiency or even tractability of its production.  
In this extended abstract, to accompany a keynote talk at the 2021 SC-Square Workshop, we survey recent work (both the author's and from the literature) on the use of Machine Learning technology to improve algorithms of interest to SC-Square.
\end{abstract}


\maketitle

\section{Introduction}
\label{SEC:Intro}

SC-Square brings together the two communities of Symbolic Computation and Satisfiability Checking, and their associated technologies of Computer Algebra Systems (CASs) and Satisfiability Modulo Theory (SMT) Solvers.  One commonality of these communities and technologies is that they value and produce exact, rather than approximate, answers to their problems.  

Machine Learning (ML) refers to statistical techniques that give computer systems the ability to \emph{learn} rules from data.  It may seem that the probabilistic nature of ML means it is of little interest to SC-Square.  However, we suggest there is great potential to use ML to uncover better strategies to optimise SC-Square algorithms and technology.

E.g., consider Buchberger's algorithm to produce the Gr\"{o}bner Basis for an ideal: a seminal result in Symbolic Computation, used as theory solver for several SMT logics.  This algorithm does not specify the order in which $S$-pairs are studied, the order is which the corresponding $S$-polynomial is reduced by the generating set, the monomial ordering to be used, and the underlying variable ordering.  Any decision for these choices allows the production of a Gr\"{o}bner Basis but each decision effects the size of the basis produced and the time taken to compute it.

ML may be able to assist with such decisions.  However, applying ML to such symbolic algebra and logic is not trivial:  there are difficult questions on how to find appropriate data; how to encode that data for ML tools; and which ML paradigm to use.  
We start this extended abstract by describing our work in Section \ref{SEC:OurWork} which attempted to use ML classification to choose the variable ordering for a Computer Algebra algorithm.  We then proceed in Section \ref{SEC:Others} to survey the literature for similar application of ML to mathematics and logic, to look for inspiration to make further progress.

\section{ML Classification for CAD Variable Ordering}
\label{SEC:OurWork}

\subsection{Cylindrical algebraic decomposition}
\label{SUBSEC:CAD}

A \emph{Cylindrical Algebraic Decomposition} (CAD) is a \emph{decomposition} of ordered $\mathbb{R}^n$ space into cells arranged \emph{cylindrically}, meaning the projections of cells are all arranged within cylinders.  The cells are (semi)-algebraic meaning each may be described by a finite sequence of polynomial constraints.  
A CAD is usually produced for a set of polynomials such that each polynomial has constant sign on each cell.  This allows us to query a finite set of sample points to understand the behaviour of the polynomials (or logical formula involving them) everywhere.  

The most important application of CAD is to perform Quantifier Elimination (QE) over the reals:  given a quantified formula, find an equivalent quantifier-free formula.  E.g., QE would transform $\exists x, ax^2 + b x + c = 0 \land a \neq 0$ into the equivalent $b^2 - 4ac \geq 0$.  Although CAD emerged in the Symbolic Computation community, since SAT is a sub-problem of QE, it can be used to tackle problems in the \verb+NRA+ and \verb+QF_NRA+ logics of the SMT-LIB.  Adaptations of the original CAD algorithm have been designed for use in SMT \cite{JdM12}, \cite{KA20}, \cite{ADEK21} and we also note the adaptation \cite{Brown2015} which is for general QE but with features inspired by Satisfiability Checking.

CAD was introduced in 1975 \cite{Collins1975} and is still an active area of research:  for a deeper introduction see, for example, the background section of \cite{EBD20}. 
 
\subsection{CAD variable ordering choice}

CAD requires a variable ordering.  For QE the ordering must match the quantification, but variables in blocks of the same quantifier and the free variables can be swapped.  So in the example above we must decompose $(x,a,b,c)$-space with $x$ last, but the other variables can be in any order.  
The ordering can have a great effect on the time / memory use of CAD, the number of cells, and even the underlying complexity \cite{BD07}.  In our example using $a \prec b \prec c$ requires 27 cells but $c \prec b \prec a$ requires 115.  
Human-designed heuristics  \cite{DSS04}, \cite{Brown2004}, \cite{BDEW13}, \cite{EBDW14} usually make the choice in implementations.  In 2014, we trained a Support Vector Machine (SVM) to choose which of these heuristic to follow \cite{HEWDPB14}.  The SVM significantly outperformed any one heuristic, identifying subclasses where each excelled.  This led to an EPSRC project and the work described in the remainder of this section.

\subsection{Results from CICM 2019}
\label{sec:d1}

The 2014 work choose between existing heuristics, in order to fix the number of classes. However, there were many problems in the dataset where none of those heuristics gave an optimal choice.  So we revisited these experiments for CICM 2019 \cite{EF19} this time allowing ML to predict the optimal ordering directly (fixing the number of variables and thus classes).  We explored a variety of ML classification methods available in Python's \texttt{sklearn} library \cite{SciKitLearn2011}: K-nearest neighbour classifiers, multi-layer perceptions, decision trees, and support vector machines.  We used the CAD implementation in Maple's Regular Chains Library \cite{CM16}.  
All the ML models outperformed the human-made heuristics for our dataset.

\subsection{Results from SC-Square 2019}

The first step to use such an ML classification model is to represent the input (in our case a set of polynomials) as a vector of floating point numbers: the \emph{features}.  In \cite{HEWDPB14} and \cite{EF19} we used measures of degree and frequency of occurrence for each variable, inspired by \cite{Brown2004}.  Then for SC-Square 2019 \cite{FE19} we developed a new feature generation procedure which evaluates combinations of basic functions (average, sign, maximum) on the degrees of the variables for individual polynomials and the system. The extra features improved the performance of all the aforementioned ML models.  Note that this feature generation procedure can be used for similar classifications where the input is a set of polynomials.

\subsection{Results from MACIS 2019}

Metrics for judging a CAD variable ordering choice should correspond to CAD runtime\footnote{A hardware independent alternative would be the number of cells produced.}.  The prior work trained ML classifiers to pick the ordering with minimal runtime for a problem, with selections deemed accurate only if that optimal ordering was chosen.  However, this meant that ML training does not distinguish between different non-optimal orderings, even though the differences are often huge.  For MACIS 2019 \cite{FE20a} we used an alternative definition of an accurate choice: one leading to a runtime within $x\%$ of the minimum.  
We then wrote a new version of the \texttt{sklearn} cross-validation procedure to select model hyper-parameters to minimise CAD runtime of the choices, rather than maximising the number of times the ordering that gives the minimal time for a problem is taken.  This improved the performance of all ML models.  Note that the new accuracy definition and procedure are suitable for any classification where we are seeking to have ML make a choice to minimise computation time.  

\subsection{Software release for ICMS 2020}

For ICMS 2020 \cite{FE20b} we presented a software pipeline that implements our work described in the previous sub-sections.  Given two datasets (training and testing) the pipeline automates:  generation of CAD runtimes for each set of polynomials under each admissible variable ordering; using the runtimes from the training dataset to select the hyper-parameters with cross-validation and tune the parameters of the ML models; and evaluating the performance of those classifiers on the testing dataset.  The pipeline could be used to pick the variable ordering for other algorithms which take sets of polynomials as input by changing the calls to Maple's CAD procedure with those of another implementation / algorithm.  The code is freely available at: \url{https://doi.org/10.5281/zenodo.3731703}.

\subsection{Success and limitations}
\label{sec:d2}

We experimented on the SMT-LIB benchmarks\footnote{http://smtlib.cs.uiowa.edu/} which are mostly real-world applications and extracted two datasets of 3 and 4 variable problems that could be tackled by CAD.  
On our 3-variable dataset human-made heuristics achieved runtimes 27\% above the minimum and ML achieved runtimes 6\% above.  So here, the ML classifiers offer close to optimal performance.  However, on the 4-variable dataset ML achieved runtime 67\% above the minimum (compared to 98\% above for human-made heuristics) and so there is room for improvement.  Of course, with 4 variables this is a much harder classification problem (24 orderings rather than 6).  

To inspire further progress we next consider related work in the literature.

\section{Inspiration from the Literature}
\label{SEC:Others}

\subsection{Other applications of ML for CAD}

The methodology of \cite{HEWDPB14} was applied later  to decide the order of sub-formulae solving in \cite{KIMA16} and whether to precondition CAD input in \cite{HEWBDP19}.  

Two more recent works with alternative methodologies are \cite{CZC20} to choose CAD variable ordering and \cite{BD20} to choose the ordering of constraints to process using the adapted CAD algorithm of \cite{Brown2015}. Both papers employed neural networks for the classification and obtained the quantity of data these need through random polynomial generation.  We note that care needs to be taken as random polynomials are known to behave quite differently to those which appear in the literature, e.g. \cite{DPSZ02} and so validation on non-random data should be encouraged.  
Both papers also took steps to tackle the large number of classes: \cite{CZC20} used an iterative greedy approach to select the ordering; while \cite{BD20} derived an ordering on the constraints from multiple binary classification on pairs.  

Applications of ML elsewhere in Computer Algebra are fairly rare\footnote{We note the early example in \cite{KUV15} which uses a Monte-Carlo tree search to find the representation of polynomials that are most efficient to evaluate.} but the following recent one may offer a blueprint for progress.

\subsection{Reinforcement learning to optimise Buchberger's algorithm}
\label{SUBSEC:GB}

Buchbergers' Algorithm to produce a Gr\"{o}bner Basis \cite{Buchberger2006} must process a list of pairs of polynomials ($S$-pairs); with that processing potentially adding further pairs to the list.  Pairs may be processed in any order, but some orders result in more pairs to study and thus more computational resources.  There exist well established strategies to make this decision (see e.g. \cite{GMNRT91}).  

In \cite{PSH20} the authors described how an Agent could be trained to make this decision using reinforcement learning: where instead of having a labelled dataset an Agent makes a decision and receives a reward that informs future decisions.  In \cite{PSH20} the Agent chooses an $S$-Pair and received a reward based on the number of polynomial additions required\footnote{Actually, the reward is based on the number of polynomial additions required to complete a full run of Buchberger's algorithm after selecting that $S$-Pair and continuing with a an existing heuristic.  The rationale for this given is to reduce variability but it seems equally compelling for allowing the Agent to judge the effects of a choice not just on the next step but on the remainder of the algorithm.  This does however greatly increase the training cost.}.  

The study ensures a constant size of polynomial by studying only binomials (so no term swell) and working in a modular coefficient field (so no coefficient swell).  They can then represent polynomials to a neural network via consistently sized exponent vectors.  Similar to our work in Section \ref{SEC:OurWork}, this allows the network to judge sparsity and degree but not the actual coefficients (to avoid over-fitting).

The experiments in \cite{PSH20} are run on separate distributions of random polynomials based on the number of variables, generators, and degree.  The Agent significantly outperformed the established strategies on such data, but the performance on real problems remains to be observed.

Most interestingly, some simple components of the Agent's strategy were observed such as a preference for pairs whose $S$-polynomials are monomials and a preference for pairs whose $S$-polynomials are low degree.  Such strategies had never been studied\footnote{Perhaps because $S$-polynomials are rarely examined before being reduced.} and when used alone outperform the established heuristics.

\subsection{ML to predict algebraic computation directly}

There has been recent work on the use of ML to predict the outcome of algebraic computations directly.  Most notably, in \cite{LC20} the authors predict the output of symbolic integration and the symbolic solutions to first and second order ordinary differential equations using neural networks\footnote{Specifically \texttt{seq2seq} models more typically used in natural language processing: for example they view integration as translating from integrands to integrals}.  Their experiments outperform various CASs, in the sense that the model predicts correct outcomes for examples where the CAS times out.  However, we note that from the viewpoint of a CAS developer the cases where the model predicts the wrong model would be more critical than the timeouts\footnote{The review \cite{Davis2019} offer some other qualifications on the claim of superiority over CASs.}.  We also note the recent preprint \cite{KTYF22} which repeats the study to make the argument that better generalisability will be achieved with a learning model based on the relative positions of mathematical symbols rather than the absolute positions.  These are very different applications of ML to the algorithm optimisation we are interested in, but lessons on how best to represent symbolic data to ML tools may well be transferable.

\subsection{ML in satisfiability checking}

An early use of ML in Satisfiability Checking was the development of the portfolio solver \textsc{SATZilla} \cite{XHHL08}.  There is rarely a single dominant SAT solver for all problems, so SATZilla uses ML to predict which solver to use for a given instance. This inspired the recent MachSMT which selects algorithms for SMT-solvers \cite{SNPNG21} using Principal Component Analysis for dimensionality reduction.

The core algorithm of Satisfiability Checking, CDCL \cite{MS99}, allows us to proceed through the exponential search space in an intelligent manner: generalising from the conflicts uncovered at a specific sample to rule out additional branches.  However, even with this conflict learning, there are decisions in the search that must be taken without guidance and poor luck can lead a search to an unproductive area, motivating for example the use of restarts.  

\noindent Thus CDCL itself has potential to be guided by ML.  For example:
\begin{itemize}
\item \cite{Wu2017} makes the choice of initial value to variable allocation to begin the search using a regression model that predicted satisfiability of formulae after fixing the values of a certain fraction of the variables. 
\item \cite{LOMTLG18} uses machine learning to determine a policy for restarts in SAT-solvers:  ML is used to predict the quality of the next learnt clause based on previously learnt clauses; restarting when the quality is predicted below a threshold.  
\item NeuroSAT \cite{SB19} can predict unsatisfiable cores (subsets of the constraints that cannot be satisfied together) to inform variable selection in the search.
\end{itemize}
The most prominent use of ML in satisfiability is probably the following one.

\subsection{Reinforcement learning for SAT-solver branching}
\label{SUBSEC:Ganesh}

The \textsc{MapleSAT} solver introduced and utilises the learning rate, the propensity for variables to generate learnt clauses, as a key metric for making decisions and the first to outperform the previously dominant VSIDS heuristic.  

In \cite{LGPC16b} they view the question of branching in SAT-solving (selecting the next variable in the search) as an optimisation problem to maximize this metric.  In particular, they apply reinforcement learning, where the learning rate informs the reward function.  Variable selection is modelled in the multi-armed bandit (MAB) framework and tackled using a well known MAB algorithm.  This led MapleSAT to victory in the annual SAT competition\footnote{https://baldur.iti.kit.edu/sat-competition-2016/index.php?cat=results}.

\subsection{ML to predict mathematical structure}

Finally, we note also the use of supervised ML to predict mathematical properties elsewhere in mathematics, where the primary motivation is the formation of new conjectures.  We refer to the survey \cite{He2022} which includes examples in algebraic geometry, representation theory, combinatorics and number theory, in which most applications are expressed as ML classification problems.

\subsection{Summary}

There is huge potential to apply ML to algorithms of interest to SC-Square.  The author's own experience in Section \ref{SEC:OurWork} shows the potential benefits. 

However, our experience using ML classification has clear limits.  Such supervised learning requires labelled datasets.  Although in theory infinite symbolic data could be manufactured, in practice it would be computationally infeasible to label all that data. A reinforcement learning approach such as for the examples in Sections \ref{SUBSEC:GB} and \ref{SUBSEC:Ganesh} looks far more promising.

Still unclear is the optimal way to represent symbolic data for ML tools, and how best to generate training data so maximise generalisation onto the problems of interest in the real world.  Such questions deserve more focussed study.

\newpage

\subsubsection*{Acknowledgements}  

The author's work surveyed in Sections \ref{sec:d1}$-$\ref{sec:d2} was joint with Dorian Florescu, and funded by EPSRC Project EP/R019622/1: \emph{Embedding Machine Learning within Quantifier Elimination Procedures}.  
The author is now supported by EPSRC Project EP/T015748/1: \emph{Pushing Back the Doubly-Exponential Wall of Cylindrical Algebraic Decomposition}.

We thank the reviewer of this paper, and the reviewers of the author's surveyed work, for useful comments.  We thank the DEWCAD Journal Club for interesting discussions on some of the other papers.


\end{document}